\documentclass{elsart1p}
\usepackage{graphicx}
\usepackage{epsfig}
\usepackage[figuresright]{rotating}
\usepackage{wrapfig}

\newcommand{\beq}{\begin{equation}}
\newcommand{\eeq}{\end{equation}}
\newcommand{\as}{\alpha_{\rm s}}

\usepackage{amssymb}
\begin{document}
\begin{frontmatter}
%
%
%
\title{Hot and dense matter: from RHIC to LHC \\ Theoretical overview}
%
%
\vskip1cm
\author{Dmitri E. Kharzeev}
\vskip0.3cm
\address{Physics Department, Brookhaven National Laboratory, Upton NY 11973-5000, USA \\ and \\ 
Department of Physics, Yale University, New Haven CT 06520-8120, USA}
\begin{abstract}
Relativistic heavy ion physics studies the phenomena that occur when a very large (in units of QCD scale $\Lambda_{\rm QCD}$) amount of energy is deposited into a large (in units of $\Lambda^{-3}_{\rm QCD}$) volume, creating an extended in space and time domain with an energy density that is large in units of $\Lambda^{4}_{\rm QCD}$.
This includes the mechanism by which the energy is deposited (likely a transformation of the colliding Lorentz-contracted  "gluon walls" into the strong longitudinal color fields); approach to thermalization; and the static and dynamical properties of the created quark-gluon plasma. Of particular interest is the fate of symmetries (e.g. chiral $SU_L(3) \times SU_R(3)$, scale, and discrete ${\cal P}$ and ${\cal CP}$ invariances) in hot and dense QCD matter. At present, the program at RHIC has entered a stage where new discoveries are enabled by high precision of the measurements; moreover, an array of new capabilities will soon be available due to the numerous and significant upgrades. Very importantly, we will soon have access to unprecedented energies of colliding ions at the LHC. In addition, future RHIC runs at low energies, FAIR at GSI and NICA at JINR will make possible the studies of  QCD matter at high baryon density. I will describe the current status of theoretical knowledge about hot QCD, and the ways in which it may be expected to improve in the near future. 

\end{abstract}
\end{frontmatter}
%
\section{Introduction}
\label{}

Relativistic heavy ion physics is placed at the intersection of three major directions of research: 
i) small $x$, high parton density QCD;  ii) non-equilibrium field theory; and 
iii) phase transitions in strongly interacting matter.
Indeed, understanding the evolution of a heavy ion collision requires a working theory 
of initial conditions, of the subsequent evolution of the produced partonic system, and 
 of the phase transition(s) to the deconfined phase.  
This (clearly incomplete and biased) overview is an attempt to capture some of the recent changes and developments in the theoretical picture of these 
phenomena which have been triggered by an intense stream of the new data from RHIC. I will also 
discuss the theoretical expectations for the physics of nuclear collisions at the new energy frontier that will soon be explored at  the LHC.  

\section{Quantum Chromo-Dynamics of strong color fields}

Before the advent of RHIC, it was widely believed that at collider 
energies the total multiplicities will become dominated by 
hard incoherent processes. The very first data from RHIC (for the overview of the data from the initial stage of RHIC experiments see \cite{Arsene:2004fa,Adcox:2004mh,Roland:2005ei,Adams:2005dq})  showed that the measured charged hadron multiplicities in $Au-Au$ collisions 
appeared much smaller than expected on the basis of incoherent superposition of hard processes.
\vskip-0.3cm
\begin{wrapfigure}{r}{0.6 \textwidth}
\noindent
\vspace{-1cm}
\includegraphics[width=8cm]{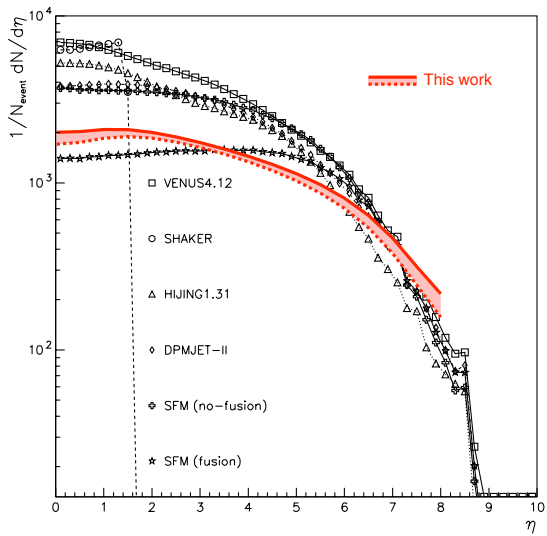}
\vspace{0.4cm}
\caption{
Comparison of the prediction for charged hadron multiplicities within the CGC-based KLN model compared to other approaches in central $PbPb$ collisions at $\sqrt{s} = 5.5$ TeV per nucleon pair at the LHC; from \cite{Kharzeev:2004if}.
}
\label{fig:kln}
\end{wrapfigure}
\vskip0.3cm
Given that any inelastic rescatterings in the final state can only 
increase the multiplicity\footnote{For statistical systems, this is due to the second law of thermodynamics}, 
we have an experimental {\it proof} of a high degree 
of coherence in multi-particle production in nuclear collisions at RHIC energies.  At high energies, all partons in the nucleus located at the same transverse coordinate contribute coherently to the scattering process, and the density of partons in the transverse plane introduces a new relevant dimensionful parameter -- the saturation momentum $Q_s$ \cite{Gribov:1984tu,McLerran:1993ni}. The density of partons in a nucleus with mass number $A$ grows as $Q_s^2 \sim A^{1/3}$, so the saturation momentum becomes quite large, $Q_s \sim 1$ GeV for a heavy nucleus at RHIC energies. One may thus hope that the corresponding coupling constant $\as(Q_s)$ is reasonably small to make weak coupling methods work in describing the dense partonic system. In this case the occupation number of gluon field modes with transverse momenta below $Q_s$ is $\sim 1/\as(Q_s)$ and large - we are dealing with a semi-classical gluon radiation field \cite{McLerran:1993ni}, or the "color glass condensate" (hereafter CGC). The small $x$ evolution thus occurs in this semi-classical background, and is described by non-linear  GLR \cite{GLR}, JIMWLK \cite{JIMWLK} and BK \cite{B,K} evolution equations.     

In a frame where both nuclei move fast (e.g. in the lab frame at RHIC and LHC), we are thus dealing with the collision of two dense "gluon walls" containing transverse color fields. Immediately after the collision, the flux of longitudinal color fields is stretched in between the remnants of the nuclei. This has long been the picture within a string model,  but in the CGC approach the created longitudinal \cite{long} ("glasma" \cite{glasma}) fields have large strength $E, B \sim Q_s^2$. These strong fields then decay into quarks and gluons creating the hot matter through a mechanism the dynamics of which is not yet entirely understood. However much progress has been made for example by solving the Yang-Mills equations to follow the real-time collision dynamics on the lattice (for concise reviews, see  \cite{Venugopalan:2008xf,Lappi:2009mp}). It is useful to have a simple analytical approach encoding the physics of parton saturation -- an example of such an approach is the KLN model \cite{Kharzeev:2000ph} that evaluates multi-particle production in nuclear collisions as a classical radiation off the colliding saturated gluon walls.  It has been found to successfully describe 
RHIC data on hadron multiplicities;  Fig.\ref{fig:kln} shows the prediction for the LHC -- again the predicted multiplicities are smaller than in most other approaches.

\section{The strongly correlated Quark-Gluon Plasma}

\subsection{Strong or weak coupling?}

The quest for the quark-gluon plasma has entered a new era  when RHIC has turned on, and will intensify at the LHC. We have already learned that the quark-gluon plasma at the temperatures accessible experimentally does not at all resemble a quasi-ideal gas of quarks and gluons, and the interactions among its constituents are very strong; these interactions for example transform a highly viscous dilute substance into what is perhaps the most perfect liquid that ever existed. Nevertheless, here I use the term "strongly correlated" instead of more familiar "strongly coupled" because at present we do not yet know whether the strong dynamical correlations that exist in the system necessarily require the coupling constant to be large, or can be reproduced by weak coupling methods through an appropriate resummation.  

At the highest temperatures $T$ accessible experimentally, the strong coupling constant is about $\as(2 \pi T) \sim 0.3$ which may 
justify an attempt to use weak coupling methods. On the other hand, the corresponding 't Hooft coupling $\lambda = g^2 N_c = 4 \pi \as N_c \sim 10$ may be large, as argued by the proponents of the strong coupling, large $N_c$ methods. Moreover, the very distinction between weakly and strongly coupled plasmas is not well-defined: we know that for different observables the  corrections (both perturbative and non-perturbative) can be very different, the situation already familiar from $T=0$ QCD. For example, current correlation functions in the quark vector channel $J_{\mu} = \bar{\psi} \gamma_{\mu} \psi$ reach the perturbative behavior already at quite small values of $Q^2 \sim 1$ GeV$^2$ whereas in the scalar gluon channel there are strong non-perturbative corrections up to $Q^2 \sim 10$ GeV$^2$; see e.g. \cite{Novikov:1981xj}. Below we will see that this situation persists also at finite temperature; the question "weakly or strongly coupled" thus is ill-defined unless we specify the observable.

\subsection{First-principle numerical calculations: lattice QCD}

\begin{figure}[htb]
\noindent
\vspace{-0.2cm}
\begin{minipage}[b]{.46\linewidth}
\includegraphics[width=7.5cm]{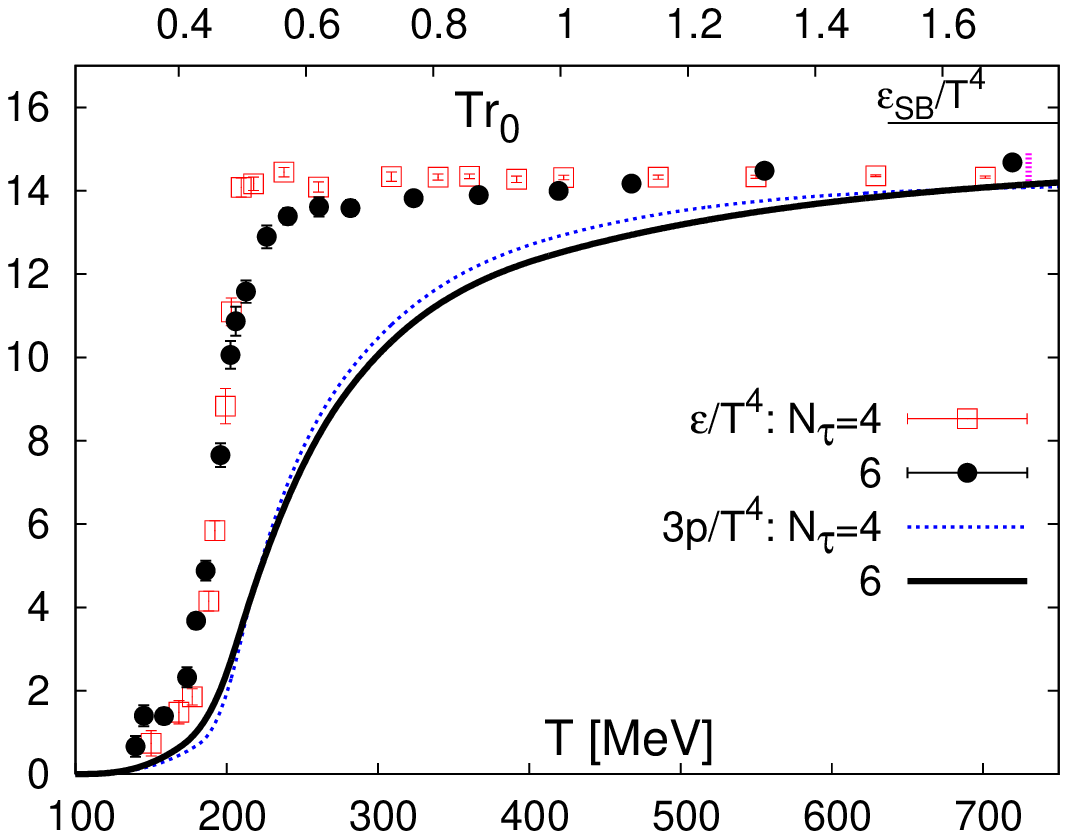}
\end{minipage}\hfill
\begin{minipage}[b]{.46\linewidth}
\includegraphics[width=7.5cm]{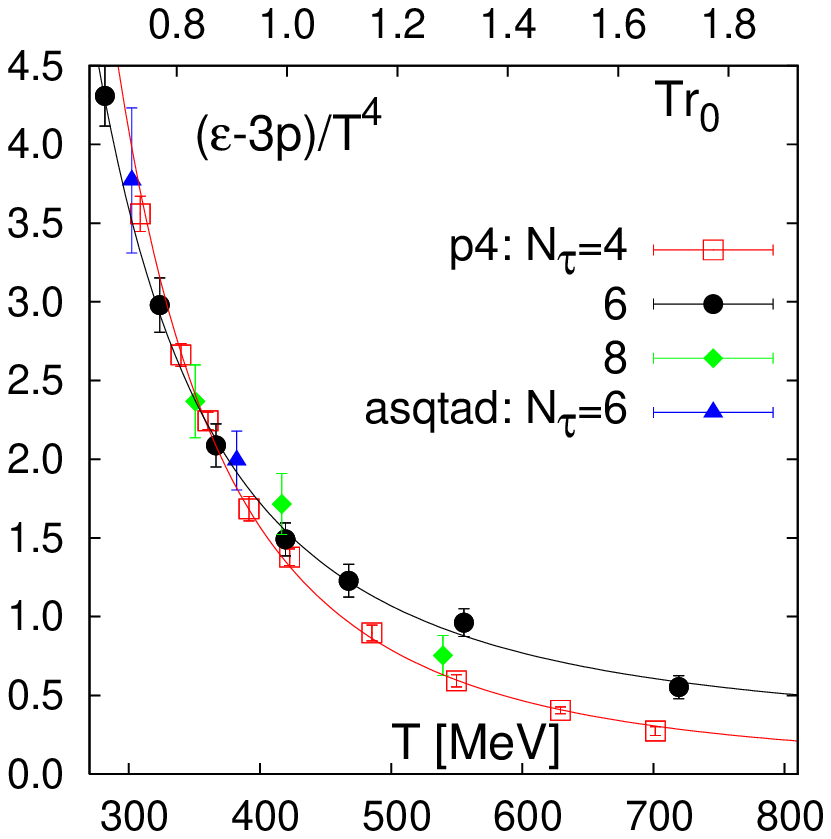}
\end{minipage}\hfill

\vspace{0.4cm}
\caption{
Left: energy density and three times the pressure of QCD matter with almost physical quark masses (two light and a heavier strange quarks) as a function of temperature; right:  the  
behavior of $(\epsilon - 3 p)/T^4$ above the deconfinement phase transition. From \cite{Cheng:2007jq}.
}
\label{fig:eos}
\end{figure}

Lattice QCD is a most valuable source of information about the properties of QCD at finite temperature. Recent rapid developments in this field (owing both to the better methods and the availability of more powerful computers) have brought us very close to the understanding of QCD thermodynamics with physical quark masses. The energy density and the pressure of the QCD plasma as a function of temperature \cite{Cheng:2007jq} is shown in Fig \ref{fig:eos}, left. To realize what temperatures are accessible at RHIC and LHC, we first evaluate the initial energy density in central $Au Au$ or $Pb Pb$ collisions using the values of saturation momentum $Q_s$ describing the multiplicities measured at RHIC and expected at the LHC, and then convert this energy density into the corresponding temperature. Following this procedure we get
\beq\label{range}
T_{RHIC}^0 \simeq 300 \div 400 \ {\rm MeV}; \ \ \ T_{LHC}^0 \simeq 500 \div 600 \ {\rm MeV}.
\eeq
One can see that already at the initial temperature achieved at RHIC the energy density is well above the critical value corresponding to the transition from hadronic gas (with $\sim N_c^0$ degrees of freedom) to the quark-gluon phase 
(with $\sim N_c^2$ degrees of freedom). One can see also that even that at the LHC the energy density and pressure do not yet reach their values expected for an ideal quark-gluon gas. Does this mean that the dynamics explored at RHIC and LHC will be qualitatively the same? 

To answer this question we need to look at the lattice data in more detail; consider for example the thermal expectation value of the trace of the energy-momentum tensor $\langle \theta_{\mu}^{\mu} \rangle /T^4 = (\epsilon - 3 p)/T^4$  that has to vanish for an ideal massless quark-gluon gas. The lattice result is shown on Fig \ref{fig:eos}, right; one can see that this quantity exhibits a steep temperature dependence in the range of temperatures given by Eq.(\ref{range}), and is significantly different at the initial temperatures reached at RHIC and the LHC. There are at least two reasons why the measurement of  $\langle \theta_{\mu}^{\mu}\rangle$ is important. First, in field theory the trace of the energy-momentum tensor is equal to the divergence of scale (or dilatational) current $s_{\mu}$ generated by the rescaling of coordinates, $\partial^{\mu} s_{\mu} = \theta_{\mu}^{\mu}$. Therefore the decrease of $(\epsilon - 3 p)/T^4$ towards higher temperatures signals the approach to an approximately scale-invariant behavior; the issue of scale invariance is crucial in deciding whether conformal theories (e.g. maximally supersymmetric ${\cal N} = 4$ Yang-Mills theory) can provide a qualitative insight into the properties of quark-gluon plasma. Second, the very character of the deviation from conformal behavior can tell us a lot about the properties of the plasma close to $T_c$. The behavior of  $(\epsilon - 3 p)/T^4$ has been fitted (see curves on Fig \ref{fig:eos}, right) by the following functional dependence  \cite{Cheng:2007jq}:
\beq
\left(\frac{\epsilon - 3 p}{T^4}\right)_{high\ T} = \frac{3}{4} b_0\ g^4 + \frac{b}{T^2} + \frac{c}{T^4},
\eeq 
where $b_0$ is the coefficient of the $\beta$ function, and $b$ and $c$ are the adjustable parameters. The first term of course arises from perturbation theory. The term proportional to $1/T^4$ stems from the zero-temperature non-perturbative physics (for example, within the bag model it is proportional to the bag constant $B$, $c \sim B$); numerically, it appears relatively small \cite{Cheng:2007jq}. The second term $\sim 1/T^2$ is the most interesting: it can be thought of as a "power correction" to the equation of state arising from an operator of dimension two\footnote{Note however that within the current precision this term cannot be unambiguously separated from a logarithmic dependence resulting from the running coupling constant.}. The physics behind this term \cite{Pisarski:2006yk,Chernodub:2008iv,Megias:2008dv} may hold the key to understanding the plasma structure close to $T_c$: such a term can arise from a contribution from magnetic monopoles or, 
in other language, from the strings that may still be present in the plasma at $T_c \leq T \leq (2 \div 3) T_c$.  The massless excitations of strings \cite{Kharzeev:2008br} with tension $\sigma$ at finite temperature contribute the $\sim \sigma/T^2$ term to the trace anomaly. Such "remnants of confinement" may be responsible for the strongly correlated nature of the quark-gluon plasma at moderate temperatures.   The role of magnetic degrees of freedom in the plasma close to $T_c$ has been emphasized e.g. in Refs.\cite{Liao:2006ry,Korthals Altes:2006gx}.

\begin{wrapfigure}{r}{0.6 \textwidth}
\noindent
\vspace{-0.8cm}
\includegraphics[width=8cm]{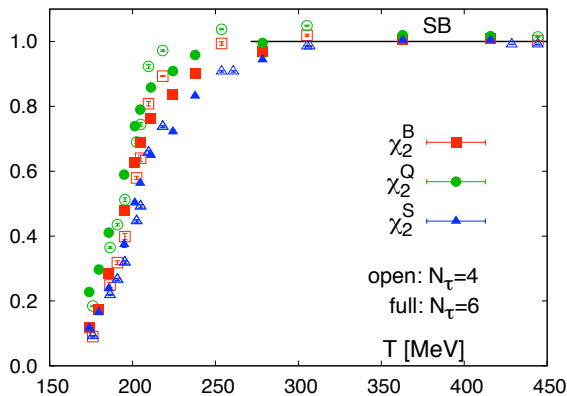}

\vspace{0.4cm}
\caption{
Quadratic fluctuations of baryon number, electric charge and strangeness, normalized to the corresponding free 
quark gas values; from \cite{Cheng:2008zh}.
}
\label{fig:fluct}
\end{wrapfigure}

\vskip0.3cm

Another important piece of information on the dynamical degrees of freedom in the plasma comes from the lattice measurements of the fluctuations of the conserved quantities: baryon number, electric charge and strangeness \cite{Cheng:2008zh}, see Fig. \ref{fig:fluct}. One can see that these fluctuations reach the values expected for an ideal quark gas already at $T \simeq 1.5\ T_c$, in sharp contrast to the case of $\langle \theta_{\mu}^{\mu} \rangle$. We are thus dealing with a very different  behavior in the vector channel (conserved quantities are the temporal components of the vector current, 
$\sim \bar{\psi} \gamma_0 \psi$) and in the scalar one, similarly to the case of zero temperature.   

\subsection{Perturbation theory and quasi-particles}  

At weak coupling $g$, quarks and gluons are the dynamical degrees of freedom in the quark-gluon plasma. The widths $\Gamma$ of quark and gluon quasi-particles are small compared to the typical thermal scale given by the temperature $T$, $\Gamma \ll T$ because they are suppressed by the powers of the coupling constant $g \ll 1$. 
Gluons acquire a dynamical Debye mass $m_D \sim g T$, and the color Coulomb potential is screened at distances 
$\sim 1/m_D$. Powerful resummation techniques have been developed to describe the thermodynamics and transport properties in this regime; a particularly useful Hard Thermal Loop method \cite{Braaten:1989mz} utilizes the hierarchy of magnetic $g^2 T$, electric $g T$ and thermal $T$ scales: $g^2 T \ll gT \ll T$. Resummation methods allow to extend the applicability of the weak coupling approach down to moderate temperatures, perhaps as low as  $T \simeq (2\div 3) T_c$ (where $T_c$ is the deconfinement transition temperature); for a review, see \cite{Blaizot:2003tw}. The parton scattering cross sections in this regime are small, $\sim g^4$, and this leads to large shear viscosity $\eta$ and small bulk viscosity $\zeta$; dimensionless ratios of these quantities to the entropy density are (up to logarithms) \cite{Arnold:2003zc,Arnold:2006fz}
\beq
{\eta \over s} \sim {1 \over g^4} \gg 1; \hskip1cm {\zeta \over s} \sim g^4 \ll 1.
\eeq 
Things become much more complicated at temperatures $T_c \leq T \leq (2 \div 3) T_c$. In this temperature range (most relevant for RHIC experiments) the 
coupling constant $g^2(2\pi T)$ is not small, and the quasi-particle description breaks down. In other words, the gluon clouds of the quark and gluon quasi-particles become so dense and the interactions between the clouds of different partons becomes so intense that the representation of this system as a superposition of individual dressed partons fails, even as a rough approximation to reality.        
 

\subsection{AdS/CFT correspondence, unparticles, and the "unplasma"}

Much of our physics intuition is based on the quasi-particle description -- most of us have a difficulty to imagine a composite object without well-defined constituents. However this kind of behavior can be encountered for example in the studies of conformal theories. Conformal theories do not possess dimensionful scales, so the spectral density of an excitation should look like a structureless mass distribution -- instead of a particle with a well-defined mass we are dealing with a broad amorphous "unparticle" \cite{Georgi}.
   This situation is somewhat reminiscent of what we expect to happen to quark-gluon plasma at strong coupling -- no well-defined quasiparticles exist, just very broad "unquarks" and "ungluons". We are thus dealing with what may be called an "unplasma" 
 of such unquark and ungluon states.     
 
 While we still do not know how to treat QCD analytically at strong coupling, there has been a breakthrough in understanding the dynamics of maximally supersymmetric conformal ${\cal N} = 4$ Yang-Mills (YM) theory based on the holographic correspondence 
 between the field theory describing behavior on $4D$ Minkowski boundary and the supergravity in the $5D$ bulk Anti-de Sitter space, supplemented by a $5D$ sphere, $AdS_5 \times S_5$ \cite{Mal9711,GKP9802,Wit9803}. The $AdS_5$ metric is unique if we want to extend Minkowski boundary space into the fifth dimension preserving the conformal invariance. A particularly useful feature of this correspondence is the duality between the strongly-coupled Yang-Mills theory and the weakly coupled gravity that can be treated by semi-classical methods. On the other hand, weakly coupled YM theory that is amenable to perturbation theory analysis is dual to strongly coupled quantum gravity. In particular, there is an intriguing relation between YM dynamics at weak coupling and  
the dynamics of gravitational collapse at strong coupling \cite{AlvarezGaume:2008qs}.
Most of the applications however deal with the strongly coupled YM theory using the duality to translate some hopelessly difficult field-theoretical problems into treatable exercises in classical gravity. 
The finite temperature $T$ is introduced on the gravity side by putting a black hole in the center of the $AdS_5$ -- note that the Anti-de Sitter metric is the solution of Einstein equations for a negative cosmological constant, and so "gravitates onto itself". The temperature $T$ and the entropy density $s$ of the YM plasma are thus  associated with a Hawking temperature of the black hole and the entropy associated with its event horizon. 
 
\subsection{Shear and bulk viscosities; plasma as a "perfect liquid"}

As a specific example, let us mention an elegant calculation of shear viscosity carried out in Ref. \cite{Kovtun:2004de}. The shear viscosity is defined by Kubo's formula through the low-frequency behavior of the correlation function of the energy-momentum tensor. 
On the Minkowski boundary, the energy-momenum tensor is an operator composed of YM fields. In gravity, the energy-momentum tensor excites gravitons, so the problem of computing the correlation function reduces to evaluating the amplitude of graviton propagation in the $AdS_5$ background.  We are interested in solving this problem in the regime of strong YM coupling, and thus of weak gravitational coupling. In this case the problem reduces to evaluating the absorption cross section $\sigma_{abs}$ of the graviton by the black hole, determined by the area of the horizon; since the entropy of the black hole is also given by the area of the horizon, one arrives at a dimensionless bound $\eta/s = 1/{4\pi}$ that is achieved at strong coupling. There is an ongoing debate on whether or not this bound is universal. Addressing bulk viscosity requires introducing deviations from the conformal behavior; however this problem may still be treatable using the gauge-gravity correspondence (see e.g. \cite{Brasoveanu:2009ky} and references therein).

\begin{figure}[htb]
\noindent
\vspace{-0.2cm}\hspace{-1cm}
\begin{minipage}[b]{.4\linewidth}
\includegraphics[width=7.5cm]{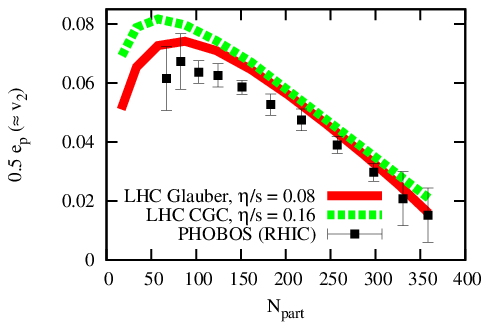}
\end{minipage}
\hspace{1.5cm}
\begin{minipage}[b]{.4\linewidth}
\includegraphics[width=7.5cm]{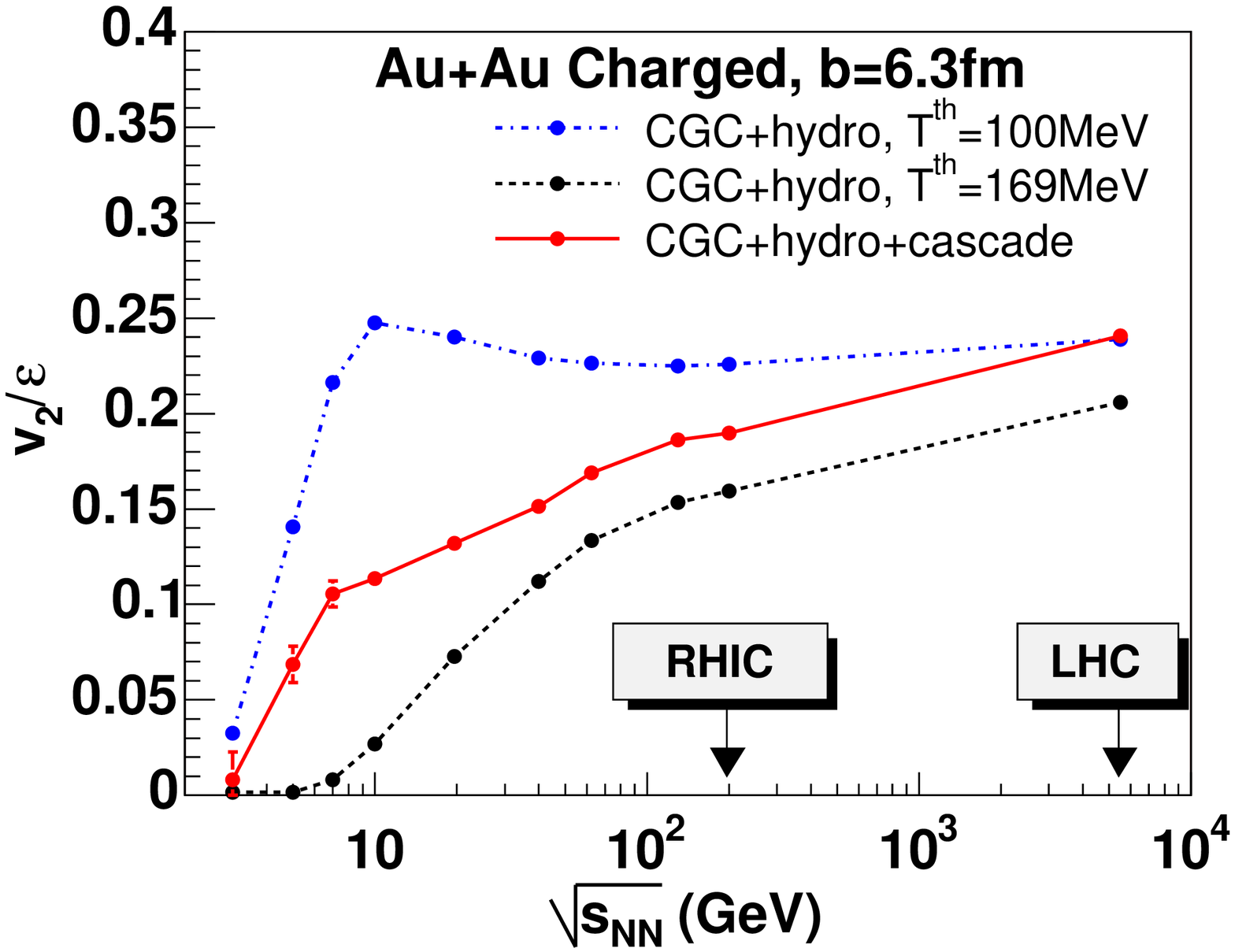}
\end{minipage}\hfill

\vspace{0.4cm}
\caption{
Left: Azimuthal anisotropy in $Pb Pb$ collisions at the LHC; from  \cite{Luzum:2009sb}.
Right: The ratio of azimuthal anisotropy $v_2$ to the eccentricity, see text for details; from \cite{Hirano:2007xd}.
}
\label{fig:hydro}
\end{figure}

In heavy ion collisions, 
 an economical way of describing the evolution of the system is 
provided by relativistic hydrodynamics, which transforms the gradients of the initial parton density into 
the momentum flow of the produced hadrons. Of particular interest is the azimuthal asymmetry of the hadron momentum distribution parameterized by the "elliptic flow" $v_2$ (for a review, see 
\cite{Voloshin:2008dg}). This quantity appears sensitive to the value of shear viscosity, as illustrated in Fig. \ref{fig:hydro}, left,  \cite{Luzum:2009sb}. It appears that the data favor small values of shear viscosity that are not far from the "perfect liquid" bound discussed above. Viscous effects contribute not only at the quark-gluon plasma stage of the evolution, but also after the hadronization; these effects can be introduced through a hadron cascade, see Fig.   \ref{fig:hydro}, right \cite{Hirano:2007xd}.

\section{Topology-induced parity violation}

\begin{wrapfigure}{r}{0.6 \textwidth}
\noindent
\vspace{-0.2cm}
\includegraphics[width=8cm]{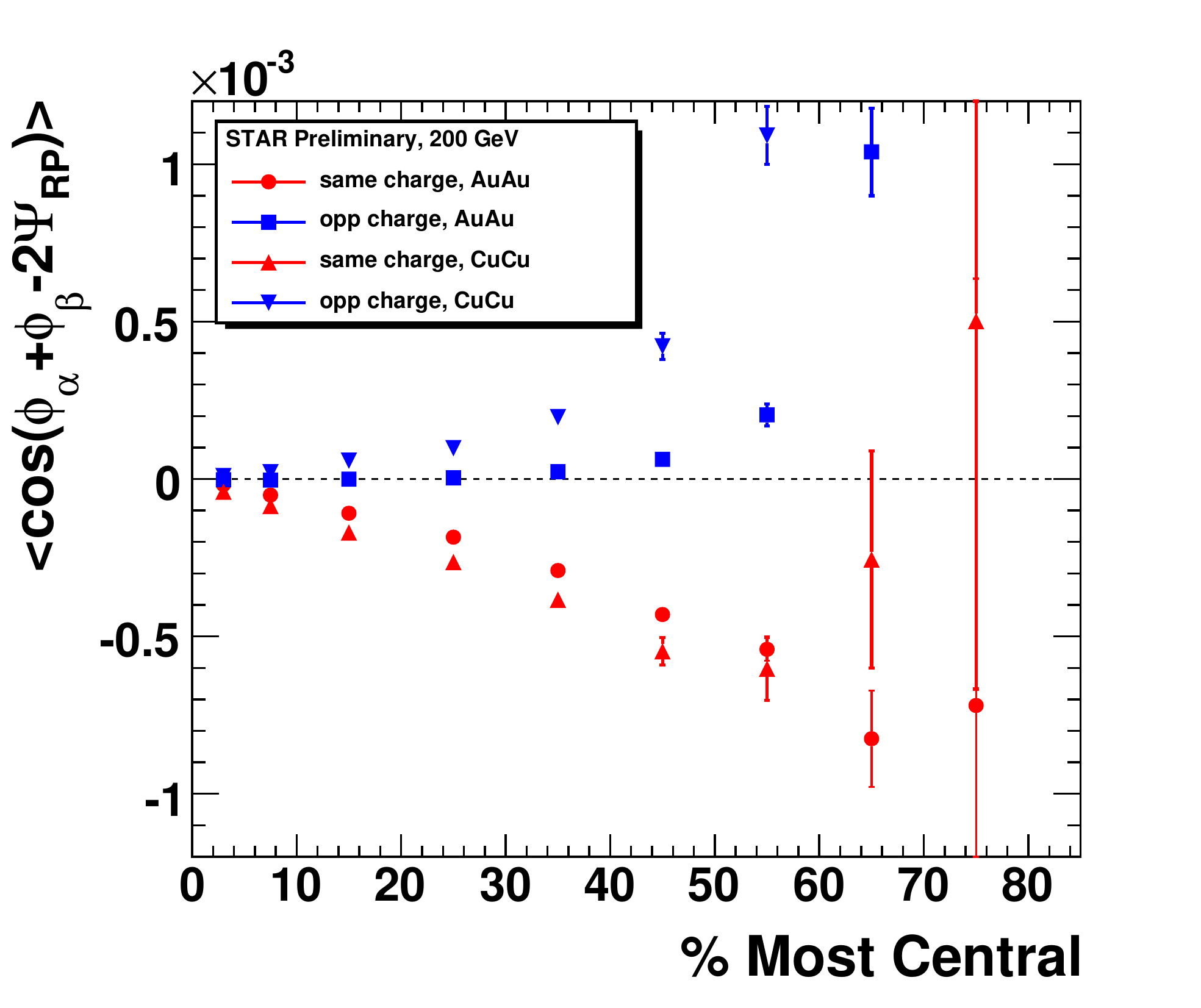}
\vspace{0.4cm}
\caption{
Correlations of charged particles with respect to the reaction plane that reveal parity violation in $AuAu$ and $CuCu$ collisions at $\sqrt{s}=200$ GeV per nucleon pair; from   \cite{Voloshin:2008jx}. 
}
\label{fig:parity}
\end{wrapfigure}

Non-Abelian nature of QCD has a very important implication for the structure of the vacuum state. Indeed, the classical Yang-Mills equations (non-Abelian analogs of Maxwell equations) possess non-trivial vacuum solutions -- instantons -- 
corresponding to the mapping 
of $SU(2)$ subgroup of the gauge symmetry $SU(3)$ onto the group of three-dimensional rotations $S_3$ \cite{Belavin:1975fg}.    
Instantons thus couple rotations in space to rotations in the space of color. 
In Euclidean space-time, instantons are static localized objects. In Minkowski space-time, instantons describe tunneling transitions between the states with different topological Chern-Simons numbers \cite{Chern:1974ft} $\nu$ of the $SU(2) \leftrightarrow S_3$ mapping \cite{'tHooft:1976fv,Jackiw:1976pf,Callan:1976je}.
At finite 
 temperature, the transition between the vacuum states with different topological numbers can occur not only through quantum tunneling, but can also be induced by a classical thermal activation process, through a "sphaleron" \cite{Klinkhamer:1984di} that induces baryon number violation in electroweak theory and helicity non-conservation in QCD \cite{McLerran:1990de}. The rate of the sphaleron transitions $\Gamma_{sph}$ is not exponentially suppressed, and is proportional to $\Gamma_{sph} \sim \alpha_{\rm s}^5 \ T^4$ (with a numerically large coefficient \cite{Moore:1997sn}). Sphalerons describe a random walk in the topological number; in a volume $V$ and time period $T$ we get the topological number $\langle \nu^2 \rangle =  \Gamma_{sph} V T$. 
 
 In off-central collisions, heavy ions possess a very large relative angular momentum and create very strong magnetic fields. 
 In this situation, the presence of topological charge was predicted \cite{Kharzeev:2004ey} to induce the charge separation with respect to the reaction plane (and thus the electric dipole moment of the produced quark-gluon plasma). Soon afterwards, an experimental observable sensitive to the effect has been proposed \cite{Voloshin:2004vk}. The first preliminary results have been reported by STAR Collaboration in Ref. \cite{Selyuzhenkov:2005xa}; Fig.\ref{fig:parity} shows the reported in \cite{Voloshin:2008jx} preliminary results for the charge asymmetries measured in $AuAu$ and $CuCu$ collisions at RHIC. 
 A detailed theory of the interplay between the chiral charge and the background magnetic field responsible for the charge separation (the "chiral magnetic effect") has been developed in Refs. \cite{Kharzeev:2007tn,Kharzeev:2007jp,Fukushima:2008xe}. The generation of chirality in the quark-gluon plasma that is responsible for the charge separation is the QCD counterpart of the generation of baryon asymmetry in the electroweak phase transition in the Early Universe \cite{Sakharov:1967dj,Kuzmin:1985mm}. 

\section{Summary}

The theory of hot and dense QCD matter is developing very fast, stimulated by the stream of high quality data from RHIC and the anticipations for the LHC. It is very hard to review all of the developments within a limited space; the most glaring omission in this talk is perhaps the theory of hard probes interacting with the QCD medium. I have no doubt that the studies of jets, high transverse momentum particles, quarkonia, heavy quarks and electromagnetic radiation that proved so important at RHIC will provide a lot more information about QCD matter at LHC.  Another omission is the physics of QCD matter at high baryon density that will be explored at low-energy runs at RHIC, FAIR and NICA - a topic of fundamental interest. 
To summarize, we have a lot to look forward to -- 
this is an exciting time to be a heavy ion physicist. 
\vskip0.1cm
I thank Itzhak Tserruya and his colleagues for their invitation to Eilat and the impeccable organization of the PANIC'08 Conference.

%
%
%

%
\end{document}